\documentclass[12pt]{article}
\usepackage[english]{babel}
\usepackage{amssymb}
\usepackage{graphicx}
\newcommand{\be}{\begin{equation}}
\newcommand{\dd}{\displaystyle}
\newcommand{\ee}{\end{equation}}
\newcommand{\bea}{\begin{eqnarray}}
\newcommand{\eea}{\end{eqnarray}}

\begin{document}
\hfill$\vcenter{
\hbox{\bf FIRENZE-DDF-392/07/02}
 }$
\begin{center}
{\Large\bf\boldmath {The linear BESS model and double\vskip.5cm
Higgs-strahlung production}}
\\ \rm \vskip1pc {\large
R. Casalbuoni and  L. Marconi}\\ \vspace{5mm} {\it{Dipartimento di
Fisica, Universit\`a di Firenze, I-50125 Firenze, Italia
\\
I.N.F.N., Sezione di Firenze, I-50125 Firenze, Italia}}
\end{center}

\begin{abstract}
In this paper we evaluate, in the context of  the linear BESS
model,  the cross-section for the double Higgs-strahlung process.
We find that, within the bounds given by the actual experimental
data, significant deviations with  respect to the SM may arise. In
the linear BESS model not only the self-couplings of the Higgs
are modified, but also the $Z$-Higgs couplings. We think that
this is a generic feature of any extension of the SM and, in our
opinion, it should be kept in mind in analyzing the future data on
the process studied here.
\end{abstract}

\section{Introduction}

During the last decade the Standard Model (SM) has been verified
with great accuracy and today there is no clear evidence  for any
deviation of the experimental data from the theoretical
expectations. However, as it is well known, there is more than one
reason to look for extensions of the SM. The main scenarios
considered for such extensions are  supersymmetry \cite{supersimm}
and technicolor \cite{technicolor}. The last type of models are
somewhat disfavored, considered the experimental data, mainly
because usually there is no natural limit in which they reduce to
the SM. On the other hand the supersymmetric models are generally
such that in the limit of heavy  supersymmetric partners of the SM
particles one gets the SM. This property is called decoupling. The
linear BESS model, although belonging naturally to a technicolor
type scenario, enjoys the decoupling property \cite{BESSlineare}.
As such it turns out to be compatible with the actual experimental
data. In the model there are six new heavy vector bosons and two
heavy scalars and it is characterized by a heavy scale of order of
a few $TeV$. Decoupling the heavy particles, as we will do in this
paper in order to get an effective lagrangian for the SM
particles, has quite different results in the low-energy physics
and in the physics associated with the standard Higgs. As far as
low-energy is concerned the effects of new physics are very soft
leading to corrections ${\cal O}(1/u^2)$ ($u$ being the heavy
scale). However there are potentially significant corrections to
the Higgs potential and to the couplings of the Higgs to the $Z$.
Therefore in this work we study  the process of double
Higgs-strahlung production where all these couplings are present
\cite{gounaris}. This process will be relevant at the future
$e^+e^-$ colliders: in $e^+e^-$ collisions the leading electroweak
processes for double Higgs production are $e^+e^- \longrightarrow
ZHH$ (double Higgs- strahlung) and $e^+e^- \longrightarrow
\bar{\nu}_e\nu_eHH$ (fusion mechanism). The second process is
relevant for C.M. energy higher than $0.6~TeV+4M_H$ \cite{BARGER}.
In this paper we will be interested at the physical capabilities
of TESLA \cite{TESLA} and therefore we will consider only the
first process,  the relevant energy at TESLA being of the order of
$500~GeV$. In this paper   we will take into account corrections
to the SM cross-section arising from the couplings of the Higgs to
the $Z$ boson since, as we shall see, these are important in the
linear BESS model. We think that this is an important point, since
it is difficult to imagine a generic model where the only
difference with respect to the SM is in the Higgs self-coupling.
In fact modifications of all these couplings with respect to the
ones of the SM appear naturally as contributions of dimension-6
operators \cite{KILIAN}.

We have organized the paper starting in Section 2 with a
presentation of the cross-section for the double Higgs-strahlung,
where we leave all the Higgs couplings arbitrary for the reasons
explained above. In Section 3 we give the expressions for the
Higgs self-coupling and for the relevant $Z$ Higgs couplings
within the linear BESS model. These couplings, as already said,
have been
  evaluated by eliminating all the heavy fields from the BESS
lagrangian obtaining an effective low-energy description of the
SM fields. Finally, in Section 4, we present our numerical
results showing that, under the constraint of compatibility with
the low-energy and LEP/SLC data, significant deviations from the
SM expectation of the  double Higgs-strahlung cross-section might
arise.

\section{Double Higgs-strahlung production}\label{sez_urto}

We give here the expression for the unpolarized differential
cross-section for the process of double Higgs-strahlung
$e^+e^-\longrightarrow ZHH$ \cite{Zerwas1} assuming  arbitrary
trilinear Higgs self-coupling, $\lambda_{HHH}$, and
couplings $\lambda_{HZZ}$ and $\lambda_{HHZZ}$ of the Higgs to
the  $Z$ boson. After integration over the angular variables one
gets:
\begin{equation}\label{sezione}
\left(\frac{d\sigma}{dx_1dx_2}\right) =
\left(\frac{d\sigma_0}{dx_1dx_2}+
 \frac{d\sigma_1}{dx_1dx_2}+\frac{d\sigma_2}{dx_1dx_2}+\frac{d\sigma_3}{dx_1dx_2}
\right) + (y_1 \leftrightarrow y_2),
\end{equation}
where:
\begin{eqnarray}
 \frac{d\sigma_0}{dx_1dx_2} & = &  \frac{\sqrt{2}G_F^3M_Z^6}{384\pi^3}\frac
    {(2c_V)^2 +(2c_A)^2}{s(1-\mu_Z)^2}\frac{2}{G_F^2M_Z^8}a^2f_0,\nonumber \\
\frac{d\sigma_1}{dx_1dx_2} & = &
\frac{\sqrt{2}G_F^3M_Z^6}{384\pi^3}\frac
    {(2c_V)^2 +(2c_A)^2}{s(1-\mu_Z)^2}\frac{\lambda_{HZZ}^4}{2G_F^2M_Z^8}
    \frac{f_1}{4\mu_Z(y_1+\mu_{HZ})^2},\nonumber \\
\frac{d\sigma_2}{dx_1dx_2} & = &
\frac{\sqrt{2}G_F^3M_Z^6}{384\pi^3}\frac
    {(2c_V)^2 +(2c_A)^2}{s(1-\mu_Z)^2}\frac{\lambda_{HZZ}^4}{2G_F^2M_Z^8}
    \frac{f_2}{4\mu_Z(y_1+\mu_{HZ})(y_2+\mu_{HZ})},\nonumber \\
\frac{d\sigma_3}{dx_1dx_2} & = &
\frac{\sqrt{2}G_F^3M_Z^6}{384\pi^3}\frac
    {(2c_V)^2 +(2c_A)^2}{s(1-\mu_Z)^2}\frac{\lambda_{HZZ}^2}{G_F^2M_Z^8}
    \frac{af_3}{2(y_1+\mu_{HZ})}. \label{2}
\end{eqnarray}
We have introduced the scaled energies of the two Higgs particles
$x_{1,2}=2E_{1,2}/\sqrt{s}$,  $x_3=2-x_1-x_2$ the scaled energy
of the  $Z$ boson and $y_i=1-x_i$. We have also introduced scaled
square masses $\mu_i=M^2_i/s$, $i=Z,H$, and
$\mu_{ij}=\mu_i-\mu_j$. We have   left unspecified the vector and
axial coupling of the $Z$ to the fermions, $c_V$ and $c_A$. The
coefficient $a$ turns out to be:
\begin{equation}\label{def_a}
a=\frac{3\lambda_{HHH}\lambda_{HZZ}}{y_3-\mu_{HZ}}+\frac{\lambda_{HZZ}^2}
         {y_1+\mu_{HZ}}+\frac{\lambda_{HZZ}^2}{y_2+\mu_{HZ}}+\lambda_{HHZZ}s,\label{3}
\end{equation}
and the coefficients $f_i$ are:
\begin{eqnarray}\label{def_f_i}
f_0 & = & \mu_Z[(y_1+y_2)^2+8\mu_z]/8,\nonumber \\
f_1 & = & (y_1-1)^2(\mu_Z-y_1)^2 -4\mu_Hy_1(y_1+y_1\mu_Z-4\mu_Z)+\nonumber \\
    &   & +\mu_Z(\mu_Z-4\mu_H)(1-4\mu_H)-\mu^2_Z, \nonumber \\
f_2 & = & [\mu_Z(y_3+\mu_Z-8\mu_H)-(1+\mu_Z)y_1y_2](1+y_3+2\mu_Z)+ \nonumber \\
    &   & +y_1y_2[y_1y_2+1+\mu_Z^2+4\mu_H(1+\mu_Z)] +4\mu_H\mu_Z(1+
          \mu_Z+4\mu_H)+\mu_Z^2, \nonumber \\
f_3 & = & y_1(y_1-1)(\mu_Z-y_1)-y_2(y_1+1)(y_1+\mu_Z)+\nonumber \\
    &   &    +2\mu_Z(1+\mu_Z-4\mu_H).
\end{eqnarray}
Each term in Eq. (\ref{2}) is easily identified by looking at the
propagators and couplings. For instance, the square of the first
term in the parameter $a$ of Eq. (\ref{3}) corresponds to the
square of the diagram which involves the trilinear Higgs coupling
$\lambda_{HHH}$. The other terms are related to sequential
Higgs-strahlung amplitudes and the four-gauge-Higgs boson
coupling.

As we have said we have left unspecified the various couplings
appearing in the differential cross-section. The reason is that
for any specified model,  all the couplings are generally
different with respect to the ones of the SM, although some of the
couplings, as for instance, $c_V$ and $c_A$ are strongly
constrained by the LEP and SLC data. In the next Section we will
consider a model for which the parameters can be chosen in such a
way to respect the experimental bounds, but nevertheless able to
produce strong deviations in the cross-section we are considering
here.
\section{The  linear BESS model}\label{fatt_vertici}

The linear BESS model \cite{BESSlineare} is an extension of the
SM involving six more vector bosons, $V_L^i$ and $V_R^i$,
$i=1,2,3$, and two more scalar particles $\rho_L$ and $\rho_R$.
The light scalar field, the usual Higgs, will be denoted here by
$\rho_U$. Furthermore it is characterized by a large mass scale,
$u$, such that for $u\to\infty$ all the new particles decouple
from the SM modes. Therefore, by appropriate choice of the scale,
it is possible to satisfy the experimental bounds  from low
energy and LEP/SLC experiments. The masses of the new particles
are of order $u$ (heavy particles), and the deviations from the
SM can be discussed in terms of the parameter expansion $v^2/u^2$
($v^2= 1/\sqrt{2} G_F$). The new particles modify the double
Higgs-strahlung cross-section since they are  coupled to the
particles of the SM (see \cite{BESSlineare}). Therefore there are
many more diagrams involved in this process than the ones coming
from the SM alone. A way of treating this problem is to  use  an
effective lagrangian describing the linear BESS model at energies
lower than the heavy scale $u$. This lagrangian  can be evaluated
solving the classical equations of motion for the heavy fields in
the low-energy limit. The full lagrangian from which the
classical equations are extracted is given in \cite{BESSlineare}.
The effective lagrangian obtained in this way has exactly the
same structure of the SM lagrangian, except that the couplings
are modified by the heavy physics. We give here the results of
this calculation concerning the couplings which are relevant in
the process we are interested in. These are nothing but the
couplings appearing in eqs. (\ref{2}) and (\ref{3}). First we
recall the expressions for the couplings in the SM:

\begin{eqnarray}\label{vert_SM}
&\lambda_{HHH_{SM}}  =  \dd{\frac{1}{2}\sqrt{\sqrt{2}G_F}M^2_H},& \nonumber \\
&\lambda_{HZZ_{SM}}  = \dd{\frac{1}{4\sqrt{\sqrt{2}G_F}}\frac{e^2}
                        {s^2_{\theta_W}c^2_{\theta_W}}},~~~
\lambda_{HHZZ_{SM}} =    \dd{\frac{1}{8}\frac{e^2}
                        {s^2_{\theta_W}c^2_{\theta_W}}},&\nonumber \\
&c_{V_{SM}}  =  -\dd{\frac{1}{2}+2s^2_{\theta_W}},~~~ c_{A_{SM}}
= -\dd{\frac{1}{2}},&
\end{eqnarray}
where $\: s_{\theta_W} \:$ is defined by:
\begin{equation}\label{COS_THETA}
s^2_{\theta_W}=\frac{1}{2}-\sqrt{\frac{1}{4}-\frac{\pi\alpha}
                  {\sqrt{2}G_FM^2_Z}}.
\end{equation}
Then we find:
\begin{eqnarray}\label{vert_BESS}
\bullet &\; HHH:& \lambda_{HHH_{SM}} \left(1-\Delta_{HHH}\epsilon
              \right), \nonumber \\
\bullet &\; HZZ:&  \lambda_{HZZ_{SM}}\left(1-
                    \Delta_{HZZ}\epsilon\right),\nonumber \\
\bullet &\; HHZZ:&
\lambda_{HHZZ_{SM}}\left(1-\Delta_{HHZZ}\epsilon
                      \right),\nonumber \\
\bullet &\; \Psi\overline{\Psi}Z:&
-\frac{e}{s_{\theta_W}c_{\theta_W}}
                         \gamma^\mu\frac{1}{2}[c_{V_{SM}} +\Delta c_V\epsilon -
                       (c_{A_{SM}} +\Delta c_A\epsilon)\gamma_5],
\end{eqnarray}
where $\epsilon = v^2/u^2$  is our expansion parameter and:
\begin{eqnarray}\label{delte}
\Delta_{HHH} & = & 3q^2,\nonumber \\
\Delta_{HZZ} & = &q^2
+\frac{e^4}{g^4_2}\frac{1-2c^2_{\theta_W}+2c^4_{
              \theta_W}}{s^4_{\theta_W}c^4_{\theta_W}}, \nonumber \\
\Delta_{HHZZ} & = &  4q^2 +\frac{e^4}{g^4_2}
       \frac{9c^4_{\theta_W}+s^4_{\theta_W}}{s^4_{\theta_W}c^4_{\theta_W}}
      +\frac{e^5}{g^5_2}\frac{8}{s_{\theta_W}c^4_{\theta_W}},\nonumber \\
\Delta c_V & = & \frac{1}{4}\frac{e^4}{g^4_2}
                 \frac{3-16c^2_{\theta_W}+18c^4_{\theta_W}-4c^6_{\theta_W}}
                      {s^4_{\theta_W}c^4_{\theta_W}(c^2_{\theta_W}-
                      s^2_{\theta_W})},\nonumber \\
\Delta c_A & = & \frac{1}{4}\frac{e^4}{g^4_2}\frac{s^4_{\theta_W}
+
           c^4_{\theta_W}}{s^4_{\theta_W}c^4_{\theta_W}},
\end{eqnarray}
where $g_2$ is
the gauge coupling of the fields $V_L$ and $V_R$, whereas  $q$
is a  parameter appearing in the scalar sector of the
BESS model \cite{BESSlineare}:
\begin{eqnarray}
V^{Higgs}        &  =  &2 \mu^2(\rho^2_L+\rho^2_R)+
                         \lambda(\rho^4_L+\rho^4_R)+2m^2\rho^2_U+
                         h\rho^4_U+ \nonumber \\
                 &      & + 2f_3\rho^2_L\rho^2_R +2f\rho^2_U(\rho^2_L
                          + \rho^2_R),
\end{eqnarray}
with $q$ defined as:
\begin{equation}
q=\frac{f}{f_3 +\lambda},
\end{equation}
We see that $q$ is the ratio among the coupling of the heavy
scalar fields to the light one and the sum of the quartic
couplings of heavy scalar fields. Therefore it is the essential
parameter (together with the ratio of the two expectation values
$v/u$) which determines the mixing of the physical Higgs field to
the heavy Higgs fields $\rho_L$ and $\rho_R$. Although  $q$ is a
free parameter of the model, we expect it to be of order one in
absence of any fine tuning of the couplings in the Higgs sector.

It is important to notice that $M_H$ and  $q$ are the only new
parameters appearing in this analysis with respect to the analysis
of the LEP data. Also notice that $q$ modifies the $HHH$ coupling
and  the couplings $HZZ$ and $HHZZ$. This is quite important
because it shows that fitting the data to a theoretical form of
the cross-section, where only the trilinear coupling of the Higgs
is left as a free parameter, might not be the right thing to be
done. In fact this example shows the possibility of correlations
among different Higgs couplings.

It will be convenient to trade the parameters $M_H$, $\epsilon$,
$g_2$ and $q$   with the set: \be
M_H,~~\frac{M_W}{M_V},~~\frac{g_{SM}}{g_2},~~q.\ee In particular:
\be \frac{M_W^2}{M_V^2}=\epsilon\frac{g_{SM}^2}{g_2^2},\ee where
$M_V$ is the mass scale of the new vector bosons. We will consider
the expansion of the cross-section around its SM value. For this
we need to keep under control the parameter $q$. From eqs.
(\ref{vert_BESS}) and (\ref{delte}) we find that the condition $q$
must satisfy is:
\begin{equation}\label{limite_teorico}
4q^2\epsilon\lesssim 1\quad\Longrightarrow \quad q\lesssim\frac{1}{2}\frac{g_{SM}}{g_2}\frac{M_V}{M_{W^\pm}}.
\end{equation}

\section{Numerical results}\label{differenza}

At  first order in the expansion parameter  $\epsilon=v^2/u^2$
the difference between the SM and BESS differential cross-section
is:
\begin{eqnarray}\label{DELTA}
&& \Delta\left(\frac{d\sigma}{dx_1dx_2}\right)  =
     \frac{\sqrt{2}G_F^3M_Z^6}{384\pi^3s(1-\mu_Z)^2}
     \frac{1}{G_F^2M_Z^8}\cdot \nonumber \\
& & \cdot\left\{2f_0\left[8\left(c_{V_{SM}} \Delta c_V+
     c_{A_{SM}}\Delta c_A\right)a^2
  -2aa_1\left((2c_{V_{SM}})^2+(2c_{A_{SM}})^2\right)\right] \right.+\nonumber\\
& & +\frac{1}{4\mu_Z(y_1+\mu_{HZ})}\left(\frac{f_1}{y_1+
          \mu_{HZ}}+\frac{f_2}{y_2+\mu_{HZ}}\right)
         \frac{\lambda_{HZZ_{SM}}^4}{2} \cdot\nonumber\\
& & \cdot\left[8\left(c_{V_{SM}} \Delta c_V+ c_{A_{SM}}\Delta
c_A\right)
      -4 \Delta\lambda_{HZZ}\left((2c_{V_{SM}})^2+(2c_{A_{SM}})^2\right)
     \right]+\nonumber \\
& & + \frac{f_3}{2(y_1+\mu_{HZ})}\lambda^2_{HZZ_{SM}}
\left.\left[8\left(c_{V_{SM}} \Delta c_V+c_{A_{SM}}\Delta
c_A\right)a
      -(a_1 + 2a\Delta\lambda_{HZZ})\left((2c_{V_{SM}})^2+\right.\right.
       \right.\nonumber\\
& &       \left.\left.\left.+(2c_{A_{SM}})^2\right)\right]\right\}
      \frac{M^2_{W^\pm}}{M^2_V}\frac{1}{(g_{SM}/g_2)^2},
\end{eqnarray}
where $a,f_i$ are defined  in Eq.~(\ref{def_a}),~(\ref{def_f_i}),
and $a_1$ is given by
\begin{eqnarray}\label{def_a_1}
a_1&=&
\frac{3\lambda_{HHH_{SM}}\lambda_{HZZ_{SM}}(\Delta\lambda_{HHH}+
    \Delta\lambda_{HZZ})}{y_3-\mu_{HZ}}+\frac{2\lambda_{HZZ_{SM}}^2
      \Delta\lambda_{HZZ} }
         {y_1+\mu_{HZ}}+\nonumber \\
& &+\frac{2\lambda_{HZZ_{SM}}^2 \Delta\lambda_{HZZ}}{y_2+\mu_{HZ}}
   +\lambda_{HHZZ_{SM}}\Delta\lambda_{HHZZ}s.
\end{eqnarray}
\begin{table}[here]
\begin{center}
\begin{tabular}{|l|c|c|c|}
\hline
$M_H(GeV)$ & $120$ & $130$ & $140$ \\
\hline \hline
$N_{HHZ}$ & $80$ & $64$ & $44$ \\
Efficiency & $0.43$ & $0.43$ & $0.49$ \\
\hline
$\delta\sigma/\sigma$ & $\pm0.17$ & $\pm0.19$ & $\pm0.23$\\
\hline
\end{tabular}
\end{center}
\caption{\small{\textit{Number of events, efficiency and
cross-section uncertainty of  double Higgs-strahlung
production with an integrated luminosity of $1000fb^{-1}$
}\rm{\cite{TESLA}}.}} \label{tab:err_TESLA}
\end{table}
The    analysis made by TESLA \cite{TESLA} (see Table 1) shows
that the estimated uncertainty in measuring the total
cross-section for the double Higgs-strahlung  is around 20\%.
Therefore, the region in the parameter space where we may expect
to find deviations to the SM within  the model considerd here is
the region complementary to the one defined by the inequality
\begin{equation}\label{varianza}
\left|\frac{\Delta\sigma}{\sigma_{SM}}\right|=
\left|\frac{\sigma_{BESS}-\sigma_{SM}}{\sigma_{SM}}\right|\le 0.2.
\end{equation}
The linear BESS parameter space gets limitations  from the LEP
and the other low energy data  only in the sector
$(M_V,g_{SM}/g_2)$ \cite{restrizione_par}. The results of this
analysis, which has been made using the data of ref.
\cite{exp_epsilon}, are given in Fig. 1, where the 95\% CL
allowed region is shown.
\begin{figure}[htb]
\begin{center}
\includegraphics[width=12cm]{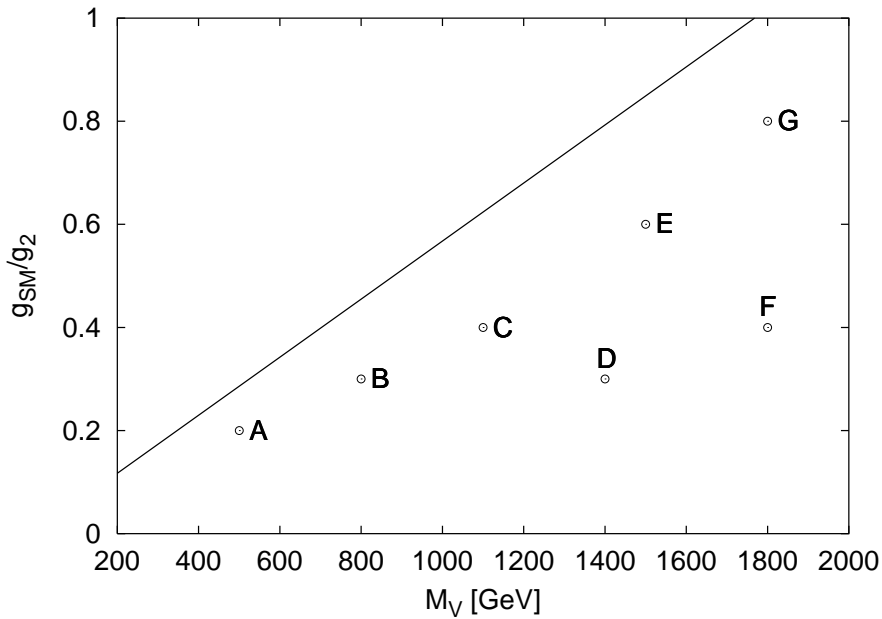}
\end{center}
\caption{\small{\textit{95\% CL allowed region in the plane
$(M_V,g_{SM}/g_2)$ from  low-energy and LEP/SLC data. We have
shown the points chosen for the present analysis.}}}
\label{fig:punti}
\end{figure}
Using these results we may eliminate the parameters
$(M_V,g_{SM}/g_2)$ from our analysis by choosing some
representative point within the allowed region. Therefore we have
selected seven points ranging in the allowed region in a range of
heavy vector masses between 500 and 1800 $GeV$. The only
remaining free parameters are now $q$ and $M_H$. However one can
see that for $110\le M_H\lesssim 200~GeV$, that is up to the
kinematical limit for producing two Higgs particles at TESLA
energy of 500 $GeV$,  the deviations do not depend on $M_H$. It
follows that in this range of values of $M_H$ the only relevant
parameter of the model  is $q$. In Fig. 2 we have plotted the
variation of the ratio $\Delta\sigma_{SM}$ with $q$ for the
points shown in Fig. 1. From this plot, for each point of Fig. 1,
we may determine the minimal value of $q$ (here denoted by
$q_{MIN}$) such that for $q>q_{MIN}$ TESLA will be able to
appreciate deviations from the SM. The numerical values of
$q_{MIN}$ obtained in Fig. 2 are reported also in Table 2 where
we have also given the limiting values $q_{lim}$ from eq.
(\ref{limite_teorico}) showing that our approximation in the
expansion of the differential cross-section is certainly valid up
to and somewhat above $q_{MIN}$.
\begin{figure}[htb]
\begin{center}
\includegraphics[width=12cm]{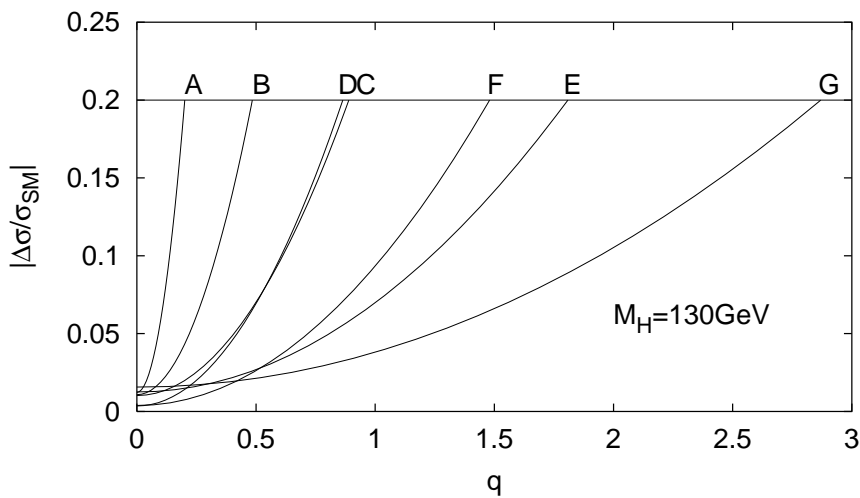}
\end{center}
\caption{\small{\textit{For $\:\sqrt{s}=500~GeV$ and
$\:M_H=130~GeV\:$, we plot
$\:\left|\Delta\sigma/\sigma_{SM}\right|\:$ as a function of  $q$
for each point chosen in the plane $\:(M_V,g_{SM}/g_2)$.}}}
\label{fig:MH130 s500}
\end{figure}
Also, since we expect $q$ to be naturally of order 1, this Table
shows that we may expect that TESLA will be able to show the
deviations given by the linear BESS model for a range of values
of $g_{SM}/g_2$ and $M_V$ not too high. Typically for
$g_{SM}/g_2\lesssim 0.5$ and $M_V\lesssim 1.5~TeV$.

In our numerical analysis we have also varied the TESLA energy
from 500 $GeV$ to 1 $TeV$ but the results are practically
insensitive to this variation of the energy.

A natural question is how radiative corrections affect the
results obtained here. The radiative corrections related to the
SM particles are the same in the linear BESS model and in the SM.
Therefore they cancel in the deviation. These corrections will be
important only when the prediction of the BESS model will be
compared with the experimental data, but of course, they are
perfectly known. The other radiative corrections related to the
exchange of heavy particles are a small fraction of the
tree-level corrections considered in this paper, and they can
safely be ignored in the present analysis.

\begin{table}[bt]
\begin{center}
\begin{tabular}{|c||c|r|c||c|}
\hline
  & $g_{SM}/g_2$ & $M_V(GeV)$ & $q_{lim}$ & $q_{MIN}$ \\
\hline \hline
A & $0.2$ & $500$ &$0.62$ & $0.20$ \\
\hline
B & $0.3$ & $800$ & $1.50$ & $0.48$\\
\hline
C & $0.4$ & $1100$ & $2.73$ & $0.88$\\
\hline
D & $0.3$ & $1400$ & $2.61$ & $0.85$\\
\hline
E & $0.6$ & $1500$ & $5.60$ & $1.80$\\
\hline
F & $0.4$ & $1800$ & $4.48$ & $1.47$\\
\hline
G & $0.8$ & $1800$ & $8.95$ & $2.86$\\
\hline
\end{tabular}
\end{center}
\caption{\small{\textit{For each point chosen in the plane $(M_V,
g_{SM}/g_2)$ we compare the value $q_{lim}$ (see text) with the
maximal value of $q$ ($q_{MIN}$) compatible with the SM.}}}
\end{table}

\section{Conclusions}

In this paper we have evaluated the cross-section for the double
Higgs-strahlung in the context of the linear BESS model. The main
property of this model is  decoupling.  This makes possible the
compatibility of the model with the actual experimental data.
Decoupling derives from the presence in the model of a heavy scale
$u$ (of the order or larger than some $TeV$) such that the
deviations (as far as the low-energy and LEP/SLC physics are
concerned) with respect to the SM are of the order $v^2/u^2$. The
model makes clear that new physics may affect the Higgs sector in
without touching the  low energy phenomenology and, more
important, has its effects in all the Higgs couplings. That is,
not only the self-coupling, but also the couplings of the Higgs to
the $Z$. This, of course, creates correlations in the amplitudes
of processes like the one studied here, and this fact  should be
taken in due care in all future analysis.

As a final comment we notice that since the model presented here
decouples at LEP energies, the only way to test it is either to
go to the $TeV$ scale where the new heavy particles should be
produced, or to look at Higgs physics, as done in this paper,
since  the new coupling $q$, involved here, is not relevant at
LEP energies. However, as far as the second possibility is
concerned, it should be stressed that presently we don't know any
way of disentangle this model from possible other models with
similar decoupling features.



\begin{thebibliography}{99}

\bibitem{supersimm} H.P. Nilles,
Phys.Rep. \textbf{110} (1984) 1;
                    H. Haber and
                    G. Kane,
Phys.Rev. \textbf{D20} (1979) 2619;
                    E. Fahri and
                    L. Susskind,
Phys.Rep. \textbf{74} (1981) 277;
                    R. Barbieri,
Riv. Nuovo Cimento \textbf{11} (1988) 1;
                    R. Barbieri,
                    F. Caravaglios and
                    M. Frigeni,
Phys.Lett. \textbf{B279} 169.

\bibitem{technicolor} S. Weinberg,
Phys.Rev. \textbf{D13} (1976) 974;
                       L. Susskind,
Phys.Rev. \textbf{D20} (1979) 2691;
                       S. Dimopoulos and
                       L. Susskind,
Nucl.Phys. \textbf{B155} (1979) 237;
                       E. Eitchen and
                       K. Lane,
Phys.Lett. \textbf{B90} (1980) 125.


\bibitem{BESSlineare} R. Casalbuoni,
                      S. De Curtis,
                      D. Dominici and
                      M. Grazzini,
Phys.Rev. \textbf{D56} (1997) 5731.

\bibitem{gounaris}
The relevance of this process was stressed by G.J. Gounaris,
D.Schildknecht and F.M. Renard, Phys. Lett. {\bf 83B} (1979) 191;
erratum {\bf 89B} (1980) 437.

\bibitem{BARGER} V. Barger,
                 T. Han,
Mod.Phys.Lett. \textbf{A Vol.5 No.9} (1990) 667.

\bibitem{TESLA} TESLA Technical Design Report, Part III
``Physics at an $\:e^+e^-\:$ Linear Collider'', TESLA Report 2001
- 23.

\bibitem{KILIAN} W. Kilian,
                 M. Kr\"{a}mer,
                 P.M. Zerwas,
Phys.Lett. \textbf{B381} (1996) 243.

\bibitem{Zerwas1} A. Djouadi,
                  W. Kilian,
                  M. Muhlleitner and
                  P.M. Zerwas,
Eur.Phys.J. {\bf C10} (1999) 27;
                  A. Djouadi,
                  H.E. Haber and
                  P.M. Zerwas,
Phys.Lett. {\bf B375} (1996) 203.



\bibitem{restrizione_par} R. Casalbuoni,
                          D. Dominici,
                          P. Chiappetta,
                          A. Deandrea,
                          S. De Curtis and
                          R. Gatto,
Phys.Rev. \textbf{D56} (1997) 2812. For a definition of the
$\epsilon$ parameters, see: G. Altarelli,
                  R. Barbieri and
                  S. Jadach,
Nucl.Phys. \textbf{B369} (1992) 3;
                  G. Altarelli,
                  R. Barbieri and
                  F. Caravaglios,
Nucl.Phys. \textbf{B405} (1993) 3;
                  G. Altarelli and
                  R. Barbieri,
Phys.Lett. \textbf{B253} (1991) 161.

\bibitem{exp_epsilon} G. Altarelli,
``The Standard Electroweak Theory and Beyond'', lectures given at
Zuoz Summer School on Phenomenology of Gauge Interactions,
CERN-TH/2000-291, hep-ph/0011078 (2000).






\end{thebibliography}
\end{document}